\documentstyle[12pt]{article}
\newcommand{\be}{\begin{equation}}
\newcommand{\ee}{\end{equation}}
\def\n{\noindent}
\catcode `\@=11
\catcode `\@=12

\title{\bf\huge A duality relation  for   fluid   spacetime}

\author{Naresh Dadhich\thanks{E-mail : nkd@iucaa.ernet.in} \\
{\sl Inter-University Centre for Astronomy \& Astrophysics,}\\
{\sl Post Bag 4, Ganeshkhind, Pune - 411 007, India.} \\
 L.K. Patel \\
 {\sl Department of Mathematics,  }\\
{\sl Gujarat University, Ahmedabad 380 009, Gujarat, India.}\\
 Ramesh Tikekar \\
{\sl Department of Mathematics, Sardar Patel University, }\\
{\sl Vallabh Vidyanagar 388 120, Gujarat, India.}}
\date{}
\begin{document}
\maketitle
\begin{abstract} 
We consider the electromagnetic resolution of gravitational field. We show that under the  duality transformation, in which active and passive electric parts of the Riemann curvature are interchanged, a fluid spacetime in comoving coordinates remains inva
riant in its character with density and pressure transforming, while energy flux and anisotropic pressure remaining unaltered. Further if fluid admits a barotropic equation of state, $p = (\gamma - 1) \rho $     where $1 \leq \gamma \leq 2$, which will tr
ansform to $p = \bigg(\frac{2 \gamma}
{3 \gamma - 2} - 1 \bigg) \rho$. Clearly the stiff fluid and dust are dual to each-other while $\rho + 3 p =0$,  will go to flat spacetime. However the radiation $(\rho - 3 p = 0)$  and the deSitter $( \rho + p = 0$ ) universes are self-dual.
\end{abstract}
\smallskip
\n PACS  Numbers  : 04.20 \\

\newpage

In analogy with the electromagnetic field, gravitational field has also been resolved into its electric and magnetic parts [1-5].It is the Weyl conformal curvature that has been resolved relative to a unit timelike vector. Recently the resolution has been
 carried through for the Riemann curvature as well [6-7].That is for the entire gravitational field including both free (Weyl) as well as matter (Ricci) parts.

We resolve the Riemann curvature into electric and magnetic parts as follows :

\be  E_{ac} = R_{abcd} u^b u^d 
\ee 

\be
\tilde E_{ac} = *R*_{abcd} u^b u^d
\ee

\be
H_{ac} = R*_{abcd} u^b u^d = H_{(ac)} - H_{[ac]}
\ee
where
\be
H_{(ac)} = C*_{abcd} u^b u^d
\ee

\be
H_{[ac]} = \frac{1}{2} \eta_{abce} R^e_d u^b u^d.
\ee
Here  $c_{abcd}$ is the Weyl conformal curvature, and   $\eta_{abcd}$   is the 4-dimensional volume element. We term  $E_{ab}$   and  ${\tilde E}_{ab} $  as active and passive electric parts which are projections of Riemann and its double dual, while magn
etic part is projection of right dual onto a unit timelike vector. Note that all these are 3-tensors as they are orthogonal to $u^a$  , the electric parts are symmetric while the magnetic part is trace-free and consists of symmetric (which is the same as 
the Weyl magnetic part) and antisymmetric parts. The electric parts account for 12 while magnetic part for the remaining 8 components of the Riemann curvature. Thus the entire gravitational field is decomposed into electromagnetic parts. The decomposition
 is however dependent upon the choice of resolving timelike vector.

The Ricci curvature can be written in terms of electromagnetic parts of the Riemann curvature as follows [7] :

\be
R^b_a = E^b_a + {\tilde E}^b_a + (E + {\tilde E}) u_a u^b -
{\tilde E } g^b_a + \frac{1}{2} \eta_{amn} H^{mn} g^b_0
\ee
where $E^a_a = E$ and  $ {\tilde E}^a_a = {\tilde E} $.  Here antisymmetric magnetic part is the measure of energy flux.
By the duality transformation we mean
\be
E_{ab} \leftrightarrow {\tilde E}_{ab}, H_{ab} \leftrightarrow H_{ab}. 
\ee
That is interchange of active and passive electric parts. Under this transformation, it is easy to see that 
\be
R_{ab} \leftrightarrow G_{ab} = R_{ab} - \frac {1} {2} R g^b_a. 
\ee
This is a reflection of the fact that contraction of Riemann is Ricci while that of its double dual is Einstein [8].

In this note we wish to examine what does this duality transformation (7) mean for energy momentum distribution of spacetime? The energy momentum tensor for a general fluid is given by
\be
T_{ab} = (\rho + p) u_a u_b - p g_{ab} + \frac {1} {2} q_{(a }u_{b)} + \pi_{ab}
\ee
where $u_au^a = 1,  \pi_{ab} = \pi_{ba}, \pi^a_a= 0 = \pi_{ab}u^b = q_au^a$.
Here  $\rho$  and  $q_a$   denote the relativistic energy density and the energy flux (heat conduction and diffusion) relative an observer comoving with 4-velocity  $u^a$ , $p$   is the isotropic pressure, and $\pi_{ab}$    is the trace-free anisotropic p
ressure (due to processes like viscosity).

In view of the Einstein equation 
\be
G_{ab} = - T_{ab}, \qquad  8 \pi G/c^2 =1 
\ee
and from eqn.(9), for a hypersurface orthogonal fluid congruence coincident with the resolving vector,  we have

\be
R^0_0 = - \frac {1} {2} (\rho + 3p), \qquad G^0_0 = - \rho
\ee
\be
R^{\alpha}_{\alpha} = - \frac {1} {2} (\rho - p )- \pi^{\alpha}_{\alpha}, G^{\alpha} _{\alpha} = p - \pi^{\alpha}_{\alpha}, (\alpha = 1, 2, 3; {\rm no \ \ summation})
\ee

\be
R_{\alpha}^0 = -q_{\alpha} u^0 = G^0_{\alpha},  \alpha = 1,2,3 
\ee

\be
R^{\alpha}_{\beta} = -\pi_{\alpha}^{\beta} = G^{\beta}_{\alpha} , \alpha \not= \beta =1, 2, 3.
\ee

First of all it is clear that distributions with trace-free energy momentum tensor (such as radiation universe with $\rho - 3 p = 0$, and Maxwell field) remain invariant under the duality transformation (7).That is they are self-dual. For a general fluid,
 the duality transformation (7) will, in view of eqns.(11-14) would imply
\be
q_a \leftrightarrow q_a, \qquad \pi^b_a \leftrightarrow \pi^b_a.
\ee
while
\be
\frac {1} {2} (\rho + 3 p) \leftrightarrow \rho, \qquad \frac {1} {2} (\rho - p) \leftrightarrow p.
\ee
Consequently if the fluid admits a barotropic equation of state, $p = (\gamma - 1) \rho$, it would transform to

\be
p = (\tilde\gamma - 1) \rho, \tilde\gamma = \frac{2 \gamma}{3 \gamma - 2} .
\ee
This means a stiff fluid $( \gamma = 2)$  is dual to dust $ ( p = 0)$, while the deSitter universe $(\rho + p = 0  )$ like the radiation universe will be self-dual. However the equation of state, $\rho + 3 p = 0$ (such as for global monopoles and global t
extures [9-10]) will go over to flat spacetime under duality transformation.

Thus we have shown that under the duality transformation (7), density and pressure of a fluid transform as given by eqn. (16) while $q_a$  and $\pi^b_a$    remain unchanged. Further if a fluid admits a barotropic equation of state with the coefficient, $\
gamma$  then it will go over to $\tilde\gamma $ . In particular the radiation and the deSitter universes are self-dual, stiff fluid and dust are dual to each-other, and the fluid with          $\rho + 3 p = 0$ will go to flat spacetime.
It may be noted that if $\gamma \leq 4/3$   (which is always so except for stiff fluid), then $\tilde \gamma \geq 4/3$, and vice-versa. One of the applications of the duality transformation could be, if one finds a solution with $\gamma \geq 4/3$,  it can
 be readily changed into $\gamma \leq 4/3 $.

Application of the duality transformation has also been considered for spherically symmetric vacuum spacetime [11]. It has been shown that the Schwarzschild particle is dual to the particle with global monopole charge, and on the other hand massless globa
l monopole and 
global texture are dual to flat spacetime. This is in agreement with the general result that fluid with the equation of state $\rho + 3 p = 0$ is dual to flat
spacetime.
\smallskip

\n {\bf Acknowledgement} \\

\n L.K. Patel and Ramesh Tikekar thank IUCAA for visit under its Associateship programme.

\end{document}